\documentclass{aastex63}
\usepackage{lineno}
\usepackage{mathrsfs} 
\received{..., }
\revised{..., }
\accepted{\today}
\submitjournal{ApJ}

\shorttitle{Reconstruction of the Total Solar Irradiance }
\shortauthors{Penza et al.}
\graphicspath{{./}{figures/}}
\begin{document}
\title{Reconstruction of the Total Solar Irradiance during the last Millenium}

\correspondingauthor{Francesco Berrilli}
\email{francesco.berrilli@roma2.infn.it}

\author[0000-0002-3948-2268]{Valentina Penza}
\affiliation{Dipartimento di Fisica, Universit\`a degli Studi di Roma Tor Vergata, Via della Ricerca Scientifica 1, Roma, 00133, Italy}

\author[0000-0002-1155-7141]{Luca Bertello}
\affil{National Solar Observatory, 3665 Discovery Dr., Boulder, CO 80303, USA}

\author[0000-0003-4898-2683]{Matteo Cantoresi}
\affiliation{Dipartimento di Fisica, Universit\`a degli Studi di Roma Tor Vergata, Via della Ricerca Scientifica 1, Roma, 00133, Italy}

\author[0000-0002-4525-9038]{Serena Criscuoli}
\affiliation{National Solar Observatory, 3665 Discovery Dr., Boulder, CO 80303, USA}

\author[0009-0000-9757-8455]{Lorenza Lucaferri}
\affiliation{Dipartimento di Fisica, Universit\`a degli Studi di Roma Tor Vergata, Via della Ricerca Scientifica 1, Roma, 00133, Italy}

\author[0000-0001-8623-5318]{Raffaele Reda}
\affiliation{Dipartimento di Fisica, Universit\`a degli Studi di Roma Tor Vergata, Via della Ricerca Scientifica 1, Roma, 00133, Italy}

\author[0000-0002-8744-4266]{Simone Ulzega}
\affiliation{Institute of Computational Life Sciences, Zurich University of Applied Sciences (ZHAW), 8820 Wädenswil, Switzerland}

\author[0000-0002-2276-3733]{Francesco Berrilli}
\affiliation{Dipartimento di Fisica, Universit\`a degli Studi di Roma Tor Vergata, Via della Ricerca Scientifica 1, Roma, 00133, Italy}




\begin{abstract}
Solar irradiance variations across various timescales, from minutes to centuries, represents a potential natural driver of past regional and global climate cold phases. To accurately assess the Sun's effect on climate, particularly during periods of exceptionally low solar activity known as grand minima, an accurate reconstruction of solar forcing is essential.
While direct measurements of Total Solar Irradiance (TSI) only began in the late 1970s with the advent of space radiometers, indirect evidence from various historical proxies suggests that the Sun's magnetic activity has undergone possible significant fluctuations over much longer timescales. Employing diverse and independent methods for TSI reconstruction is essential to gaining a comprehensive understanding of this issue.
This study employs a semi-empirical model to reconstruct TSI over the past millennium. Our approach uses an estimated open solar magnetic field ($F_{o}$), derived from cosmogenic isotope data, as a proxy for solar activity. 
We reconstruct the cyclic variations of TSI, due to the solar surface magnetic features, by correlating $F_{o}$ with the parameter of active region functional form. Instead, we obtain the long-term TSI trend by applying the Empirical Mode Decomposition (EMD) algorithm to the reconstructed $F_{o}$ to filter out the 11-year and 22-year solar variability.  
We prepare a reconstructed TSI record, spanning 971 to 2020 CE. The estimated departure from modern TSI values occurred during the Sp\"{o}rer Minimum (around 1400 CE), with a decrease of approximately 2.3 $W m^{-2}$. A slightly smaller decline of 2.2 $W m^{-2}$ is reported during the Maunder Minimum, between 1645 and 1715 CE.

\end{abstract}

\section{Introduction} 
\label{sec:intro}

Solar radiation is the most significant energy contributor to the Earth's energy budget \citep{kren2017,lecuyer2015} and is a crucial external factor influencing the global climate \citep{jungclaus2017, IPCC}. Similarly, stellar irradiance and its variability affect the atmosphere of exoplanets and their habitability \citep[e.g.][]{Galuzzo2021,linsky2017,modi2023}, and affect our ability to detect exoplanets \citep[e.g.][]{Galuzzo2021,rackam2023}.

Total Solar Irradiance (TSI) is the amount of solar radiative energy integrated over the entire spectrum measured at a distance of 1 AU. Prior to the space age, which enabled precise measurements of solar irradiance outside the Earth's atmosphere, TSI was considered unchanging over time, to the extent that it was defined as \emph{solar constant}.
Highly accurate measurements from space commenced in the late 70s and have allowed assessing TSI variations over timescales ranging from minutes to decades \citep[e.g.][]{Foukal1988,Lean1995,Solanki1999,Shapiro2011r,kopp2016}. In particular, the periodicity of TSI variations over the 11-year solar cycle has been assessed. These variations, in phase with the solar cycle, can be quantified at approximately $0.1 \% $ from minimum to maximum \citep[e.g.][]{kopp2016}.

Although based on different methodologies, these models are predicated on the empirical evidence that the TSI is modulated by photospheric magnetic field concentrations. Sunspots provide a negative contribution, while faculae and network provide positive contributions \citep[e.g.][]{Steinegger1996,Berrilli1999,Ermolli2003}. During the 11-year cycle, the contribution from faculae exceeds that of sunspots, resulting in a positive correlation between TSI and other activity indices, as for instance the sunspot number, at this temporal scale.

Space observations have been only gathering data for roughly 40 years. Variations on centennial scales are clearly visible in the trend of maxima of the Sunspot Number (SSN) dataset, which represents the direct solar observable covering the longest historical period \citep[e.g.][and references therein]{Gleissberg1939,Arlt2020}. In particular, the SSN observations over the past 400 yr have revealed the presence of periods of grand maxima and minima of solar activity \citep[e.g.][]{Stuiver1998,Vonmoos2006,Usoskin2007,Abreu2008,Vecchio2017}. Among the grand minima, the Maunder Minimum, occurring in the latter half of the 17th century, stands out as the most widely studied.  
The estimate of irradiance variations from timescales of decades to millennia is an important input to global Earth's climate models \citep[see e.g.,][]{lockwood2012, solanki2013,Bordi2015,matthes2017,liu2023}. 
A variety of models have been proposed in the literature that aim at reproducing solar irradiance variability over different periods, ranging from months \citep[e.g.][]{Willson1981,Oster1982,Sofia1982,Foukal1988}, to years \citep[e.g.][]{yeo2017, yeo2017JGRA, penza2003,lean2020}, to centuries  \citep[e.g.][]{preminger2006,tapping2007,coddington2016,yeo2014, wu2018,penza2022} and millennia \citep[e.g.][]{lean2018, abdullah2021}. Recent reviews of TSI reconstructions are provide in \citet{faurobert2019, petrie2021, kopp2021, chatzistergos23}. Here. it is important to recognize that irradiance reconstructions before the twentieth century typically depend on proxies (sunspot number or sunspot groups and radioisotopes). Models vary from using correlations between irradiance and proxies based on modern observations \citep{steinhilber2009, lean2018}, to employing models of varying complexity to determine the distribution of magnetic fields across the solar disk \citep[e.g.][]{wang2005, wu2018}.

This work is the natural continuation of the approach proposed in \citet{penza2022}. In that paper, the modulation on centennial timescales (from 1513 EC to the present) was derived from the decomposition of the Solar Modulation Potential $\phi$ \citep{muscheler2016, brehm2021} in different time-scales components derived with an Empirical Mode Decomposition algorithm. 

The solar modulation potential $\phi$ represents the average energy loss of a cosmic ray from
the heliopause to the Earth, due to the interaction with the heliosphere \citep[see e.g.,][]{Gleeson1968,Caballero-Lopez2004}. Because the interplanetary physical and magnetic conditions change
with the level of solar magnetism, $\phi$ is a proxy of solar magnetic activity. The modulation potential is
typically calculated using data from multiple Neutron Monitors  \citep[see e.g.,][]{simpson}, following a method initially proposed by \cite{Usoskin2005}. For a detailed explanation, please refer to the paper by \citet{Usoskin2005}.

In order to extend the TSI reconstruction backward by a thousand years and provide detailed information on individual cycles, we apply here the same approach to the Open Solar Magnetic Field ($F_{o}$). $F_{o}$ and $\phi$ are strongly correlated physical quantities, as the latter is a parameterization of the cosmic rays intensity, which in turn is modulated by the heliospheric magnetic field.

The relationship between the cosmic ray intensity and $F_{o}$ has been investigated and established by several works \citep[see e.g.,][]{Caballero-Lopez2004, McCracken2007}, and in \citet{steinhilber2010} an analytical power-law relationship between $F_{o}$ and $\phi$ is derived.
Here we use the $F_{o}$ dataset reconstructed by \citet{usoskin2021} with  annual cadence for the period 971 - 1899 EC, based on cosmic ray flux data assessed from cosmogenic-isotope $^{14}C$ measurements in tree rings \citep{brehm2021}, and we extend this dataset for the period the 1900-2020 EC by using an empirical relation between $F_{o}$ and $\phi$, similarly to that made in \citet{steinhilber2009}.
Moreover, in order to estimate the 11-year variations, we use updated composites of sunspot and plage area coverages \citep{mandal2020, chatzistergos20}, that cover the temporal periods 1874-2023 EC and 1893-2023 EC, respectively.

The paper is organized as follows: in Sect. \ref{sec2}, we provide a description of the $F_{o}$ dataset; in Sect. \ref{sec_emd}, we describe the procedure used to extract the long-term modulation function; Sect. \ref{sec_plage_sunspot} describes the technique used to reconstruct the sunspot and plage coverages; the Total Solar Irradiance reconstruction is shown and described in Sect. \ref{sec_tsi}; finally, in Sect. \ref{sec_discussion},
we present a discussion of the results followed by a concise summary.

\section{Open solar magnetic flux dataset} \label{sec2}

\begin{figure}[h]
\centering
\includegraphics[scale=0.7]{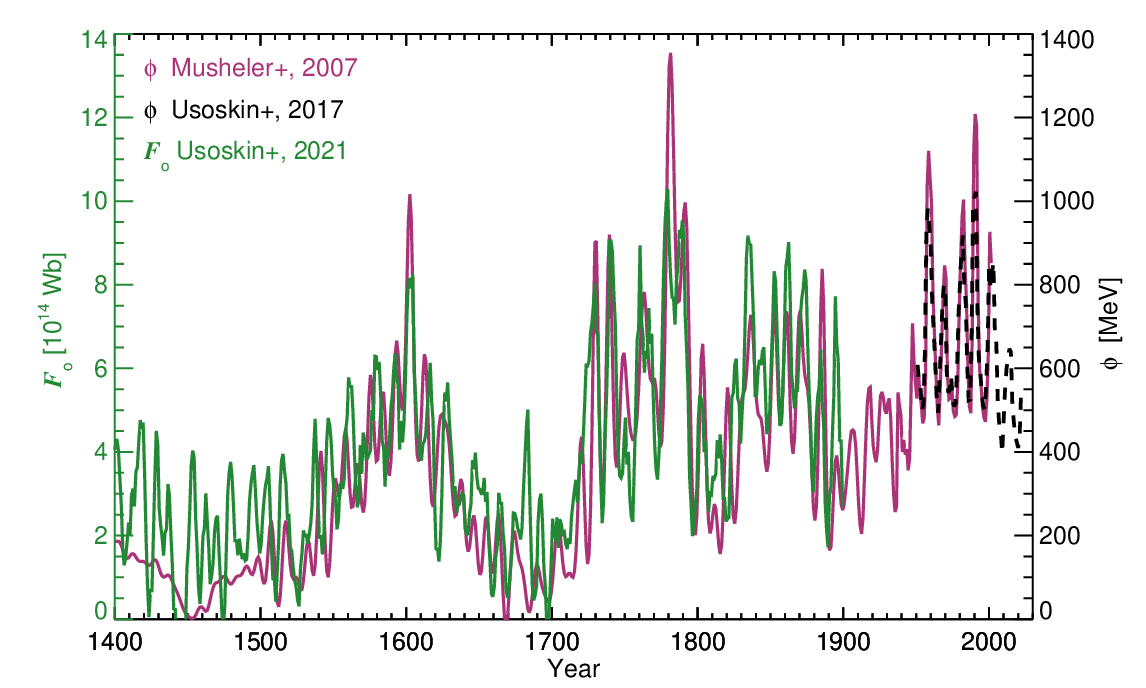}
\caption{Comparison of the annual values of open solar magnetic flux $F_o$, computed by  \citet{usoskin2021} (green line), with the solar modulation potential $\phi$ values computed by \citet{muscheler2007} (purple line) and \citet{usoskin2017b} (black dashed line).}
\label{Fopen_phi}
\end{figure}

\begin{figure}[h]
\centering
\includegraphics[scale=0.7]{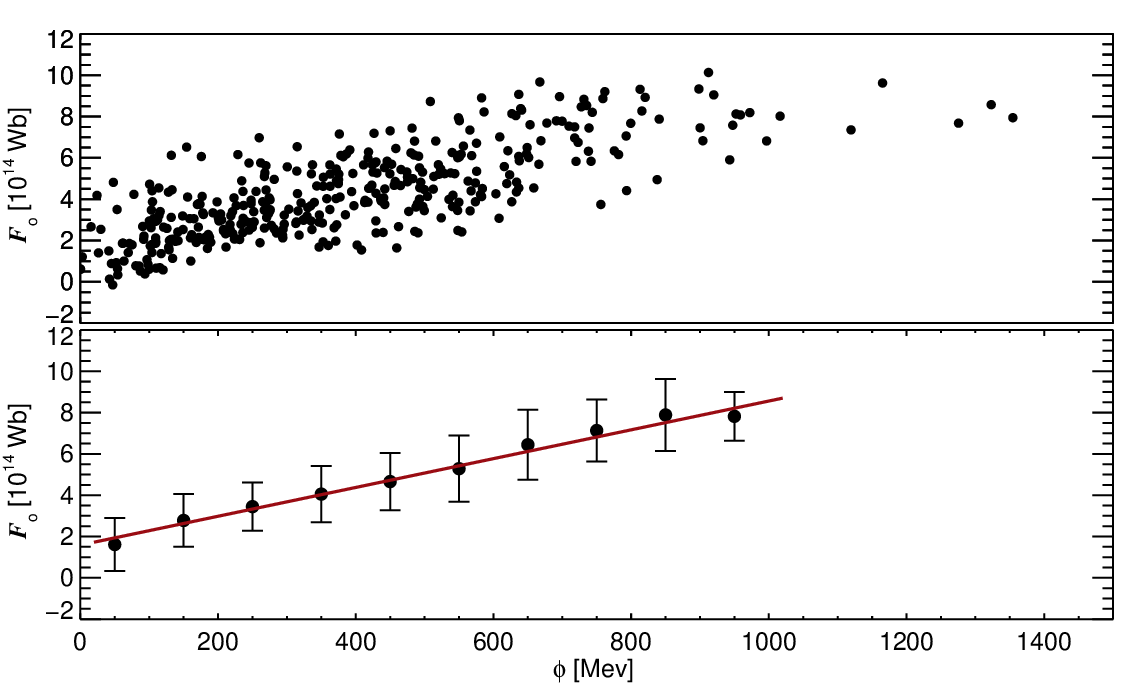}
\caption{Top panel: The relationship between the annual values of open solar magnetic flux $F_o$ from \citet{usoskin2021} and the solar modulation potential $\phi$ from \citet{muscheler2007} for the years 1513 to 1899. Bottom panel: The same relationship after the data were divided into Gaussian bins
with a FWHM of 100 Mev. The error bars represent the standard deviation of the weighted mean values. The red line represents the linear regression
computed for $\phi$ value less than 1000 Mev (Eq. 1 in the text). The Pearson's correlation coefficient is 0.996, statistically significant at $p < 0.01$}
\label{Fopen_vs_phi}
\end{figure}

We use the dataset of $F_{o}$ as reconstructed by \citet{usoskin2021} \footnote{\url{http://cdsarc.u-strasbg.fr/viz-bin/cat?J/A+A/649/A141}.}.
This reconstruction contains data from 971 to 1899 CE with annual cadence. In order to extend the data up to the present, we exploit a simple relation between the values of $F_{o}$ and the values of the solar modulation potential $\phi$ ($F_{o}$ and $\phi$ are shown in Fig. \ref{Fopen_phi}).
This relation is provided by the correlation between the values of $F_{o}$ and $\phi$  by \citet{muscheler2007} in the period 1513-1899 CE, that is shown in Fig.\ref{Fopen_vs_phi}.
This specific time interval was chosen for comparison because, prior to this period, it is not possible to distinguish the 11-year cycle variations in the $\phi$ dataset. \citet{muscheler2007} attribute it to the different time cadence in the $^{14}C$ record. 
Furthermore, due to a limited number of points, the data bove $\phi =$ 1000 MeV may not be statistically significant.
Therefore, we consider the parameters of a linear fit obtained for $\phi <$ 1000 MeV. This choice is supported by the fact that for the period during which we will use this relation (after 1900 EC), the composite $\phi$ presents a maximum value of 1022 MeV occurring in the 1991 EC. The relation is the following:  

%%%%%%%%%%%%%%%%%%%%%%%%
\begin{equation}
\label{phi_vs_Fo}
F_{o} = (0.008 \pm 0.002) \,\, \phi  + (1.4 \pm 0.9)  ~~~ 10^{14} Wb  .
\label{eqF0}
\end{equation}
%%%%%%%%%%%%%%%%%%%%%%%%%

The Pearson's correlation coefficient for this relation is 0.996, statistically significant at $p < 0.01$.

We apply Eq. \ref{eqF0} to the  $\phi$ composite for the period 1513-2020 CE, obtained from the combination of two datasets: \citet{muscheler2007} for 1900-1949 CE and \citet{usoskin2017b}  \footnote{\url{https://cosmicrays.oulu.fi/phi/phi.html}}, for 1950-2020 CE. 
%%%%%%%%%%%%%%%%%%%%%%%%% 
%%%%%%%%%%%%%%%%%%%%%%%%%%%%%%%%%%%%%%55

%%%%%%%%%%%%%%%%%%%%%%%%% 
%%%%%%%%%%%%%%%%%%%%%%%%%%%%%%%%%%%%%%

\begin{figure}[h]
\centering
\includegraphics[scale=0.7]{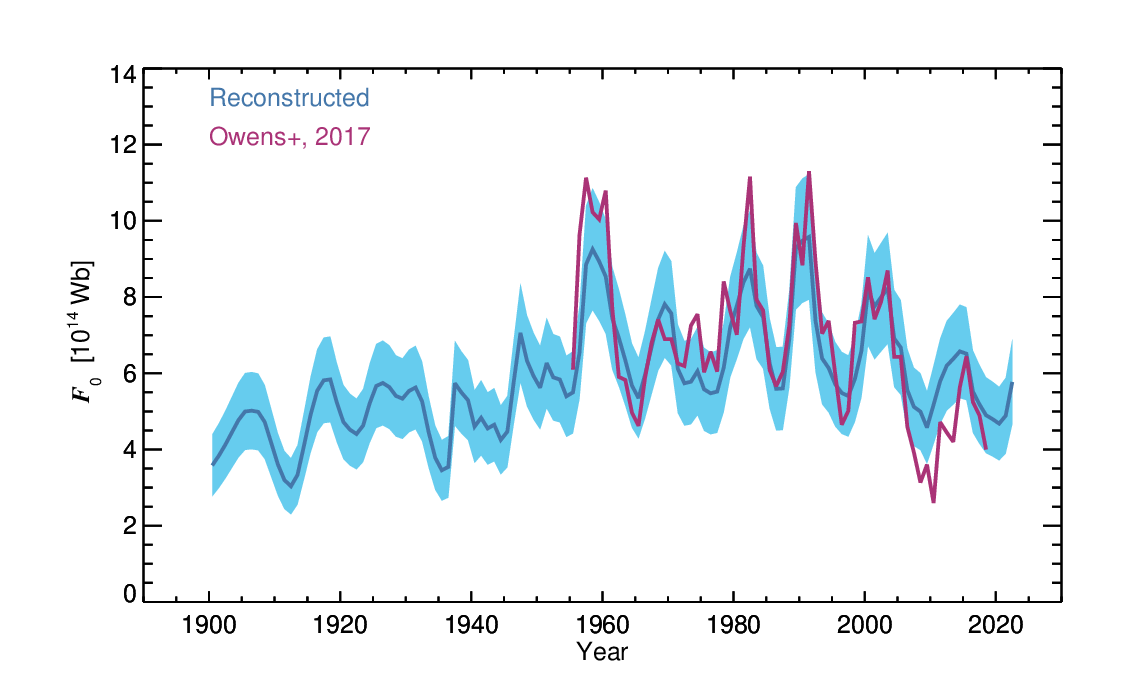}
\caption{Reconstruction of the open solar magnetic flux $F_o$ for the last century is depicted in blue. The light blue region illustrates the uncertainty range in the reconstructed time series. For comparison, data from \citet{owens2017} spanning the time interval 1950-2022 are shown in red.} 
\label{Fopen_REC}
\end{figure}

The $F_{o}$ reconstruction  for the post-1900 period, based on the linear relation provided above, is shown in Fig. \ref{Fopen_REC} where the observational data of \citet{owens2017} are reported for comparison. The confidence interval is computed by propagation of the errors in the fit coefficients in Eq. \ref{eqF0}. We note that the reconstruction is compatible with experimental data within the error bars for almost all the years, except for a slight overestimation for the years between 2008 and 2014, corresponding to the minimum between the Cycle 23 and 24. \citet{owens2017} highlighted possible uncertainties in their dataset during periods of minimal activity. 
%%%%%%%%%%%%%%%%%%%%%%%%% 
%%%%%%%%%%%%%%%%%%%%%%%%%%%%%%%%%%%%%%

We use the composite of $F_{o}$ both to reconstruct active region coverages and to obtain a long-term modulation function ($F_{LT}(t)$), which is not present in the sunspot signal, to be applied to reconstruct the quiet network component. The latter is computed using a method similar to the one proposed in \citet{penza2022}: we decompose $F_{o}$ by employing the EMD; we consider the Intrinsic Mode Functions (IMFs) corresponding to the components not present in sunspots and the monotonic residual signal; finally we properly standardize the data.

%%%%%%%%%%%%%%%%%%%%%%%%% 
\begin{figure}[h]
\centering
\includegraphics[width=\textwidth]{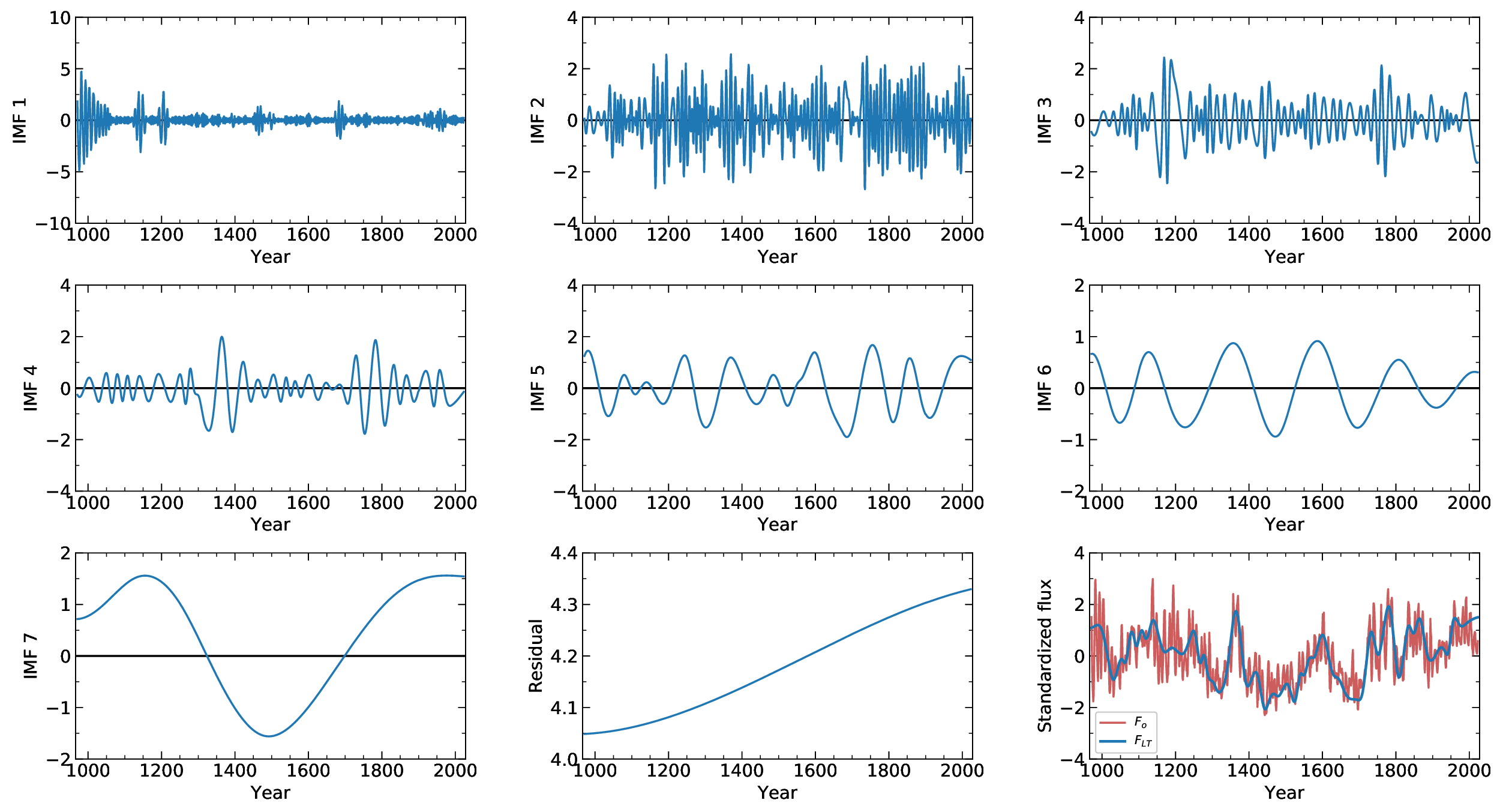}
\caption{The IMFs extracted with the Empirical Mode Decomposition applied to the reconstruction of $F_{o}$. The panels in the first two rows, together with the leftmost in the third row show the successive order IMFs. The center panel in the third row shows the residual signal of the decomposition. The modulation function ($F_{LT}$), resulting from the standardized sum of the IMFs 4-7 and the residual, is shown in the bottom right panel, along with the standardized $F_{o}$.}
\label{IMF}
\end{figure}
%%%%%%%%%%%%%%%%%%%%%%%%%%%%%%%%%%%%%%

\section{Empirical Mode Decomposition and Hilbert Spectral Analysis}\label{sec_emd}
The Empirical Mode Decomposition (EMD) \citep{Huang1998} is a data-driven decomposition technique tailored to non-linear and non-stationary signals, which makes possible to capture oscillatory modes with variable frequencies, enabling the examination of signal components across different scales. This outstanding feature of the EMD has made it a widely used technique in several fields of physics, such as space and solar physics. In particular, within the latter it has been used to study both short-term \citep{Kolotkov2015a} and long-term periodicities of solar activity \citep{Li2007, Vecchio2019}, to extract the Space Climate variability in solar UV emission \citep{lovric2017} and solar wind properties \citep{Reda2024}, to reveal the presence of quasi-periodic pulsations in solar flares \citep{Nakariakov2010, Kolotkov2015b}, as well as to identify the propagation of waves in the solar atmosphere \citep{Terradas2004,Stangalini2014,Jefferies2019}.
One of its key features is that the decomposition basis is not a-priori prescribed, in contrast to methods like Fourier analysis which relies on sinusoidal basis, or wavelet analysis which utilizes a predefined mother wavelet. Indeed, the EMD strength lies in the adaptability of the decomposition basis, which is simply derived from the inherent properties of the signal under analysis via an iterative procedure known as \textit{sifting process} \citep[see e.g.,][]{Rilling2003}. This procedure, which is based on interpolation by means of cubic splines, is such that each of the extracted modes have a zero-average mean envelope. The output of the sifting process is a set of Intrinsic Mode Functions (IMFs), each one representing a mode of oscillation embedded in the starting signal. To be defined as such, an IMF must have the same number of extrema and zero crossings, or differ by at most one. Overall, the EMD enables to express any generic time-series $s(t)$ as a finite sum of $n$ time-dependent modes of oscillation (i.e., the IMFs) plus a residual term $R(t)$ that captures the time-trend of the signal.

\begin{equation}
    s(t) = \sum_{j=1}^{n} \mathrm{IMF}_{j}(t) + R(t)
\end{equation}

As previously mentioned, each IMF represents a proper mode of oscillation of the signal on a particular characteristic time scale. 
The EMD approach applied to the reconstruction of $F_o$ results in 7 IMFs and a monotonic residual, which are shown in the subplots of Figure \ref{IMF}. The corresponding characteristic time scales, as derived via the Hilbert Spectral Analysis, are: $\tau_{1} \simeq 3.6$ yrs; $\tau_{2} \simeq 10.5$ yrs; $\tau_{3} \simeq 21.6$ yrs; $\tau_{4} \simeq 45$ yrs; $\tau_{5} \simeq 115$ yrs; $\tau_{6} \simeq 211$ yrs; $\tau_{7} \simeq 860$ yrs. The estimation of the characteristic scale allows us to associate the majority of the components to known periodicities of the solar activity. The first component (IMF 1) likely corresponds to Quasi-Biennial Oscillations (QBOs), that are associated with the Gnevyshev gap \citep[e.g.,][]{Gnevyshev1967}, which in turn is responsible for the double peak of the solar cycle \citep[see e.g.,][]{Bazilevskaya2014}; the IMF 2 and the IMF 3 are clearly associated to the Schwabe cycle \citep[e.g.,][]{Schwabe1844} and the Hale cycle \citep[e.g.,][]{Hale1919}, with IMF 3 possibly also reflecting a contribution from the Gnevyshev-Ohl rule \citep{Gnevyshev1948}; the fourth component (IMF 4) may represent an harmonic of the Hale cycle; the IMF 5 shows a periodicity associated to the Gleissberg cycle \citep[e.g.,][]{Gleissberg1939}; the IMF 6 has a time scale compatible with the Suess-de Vries cycle \citep[e.g.,][]{Suess1980}; finally, the last component (IMF 7) may be linked to a lesser-known periodicity known as Eddy cycle \citep[e.g.,][]{Steinhilber2012, usoskin2017}.
At this stage, since our aim is to extract only the long-term modulation from $F_{o}$, we filter out the contribution of the components with $\tau \lesssim 22$ yrs (i.e., IMF 1, IMF 2 and IMF 3). The resulting signal, later standardized, is shown in the bottom-right subplot of Figure \ref{IMF}. Such signal, namely $F_{LT}$, constitutes the long-term modulation in the subsequent reconstructions.

The long-term modulation derived from $F_{o}$ is utilized to reconstruct the impact of the magnetic fields of the quiet Sun on Total Solar Irradiance variability.
In the subsequent section, we will delve into the exploration of the contribution of magnetic fields from active regions, such as sunspots and faculae, to the TSI.

%--------------------------SECTION--------------------

\section{Plage and sunspot coverages reconstruction} 
\label{sec_plage_sunspot}

The irradiance variation on shorter timescales, up to the decade, are mainly modulated by the presence of bright and dark magnetic regions and by the disk surface fraction covered by them. In order to reconstruct this component we use the same approach as in \citet{penza2021,penza2022}. 
We utilize the functional form presented in \citet{Volobuev} to mathematically represent individual sunspot cycles:

%%%%%%%%%%%%%%%%%%%%%
\begin{equation}
\label{cycle_form}
x_{k}(t) =  \left(\frac{t - T0_{k}}{Ts_{k}}\right)^{2} \textrm{exp}
\left[-\left(\frac{t - T0_{k}}{Td_{k}}\right)^{2} \right]
\quad T0_{k} < t < T0_{k} + \tau_{k}
\end{equation}
%%%%%%%%%%%%%%%%%%%%%%%

with $T0_k$ and $\tau_k$ denoting the start year and duration of cycle $k$, respectively, and $x_k(t)$ refers to a solar activity proxy, such as
sunspot area coverage or plage area coverage. The free parameters
$Td_k$ and $Ts_k$, both measured in years, characterize the overall shape of cycle $k$. 
Within the time interval of cycle $k$, the function $x_k(t)$ reaches its maximum, $(Td_k/Ts_k)^2\exp(-1)$, at the time $t = T0_k + Td_K$. 
We define the parameter $P_k$ as the ratio $(Td_k/Ts_k)^2$, which serves as an indicator of the peak intensity for cycle $k$.

The relationship between $Ts_{k}$ and $Td_{k}$ is known: cycles with large amplitude (smaller $Ts_{k}$) present a shorter time of rising to maximum (shorter $Td_{k}$). This is known as the Waldmeier rule.

As the dataset of the sunspot area coverage \citep{mandal2020} is the same as in \citet{penza2022}, we use here the same $Td_{k}$ and $Ts_{k}$ (given in units of year) with their mutual relation:
%%%%%%%%%%%%%%%%%%%%%%%%
\begin{equation}
\label{TsvsTd_spot}
Td_{k}^{spot} = s_1 Ts_{k}^{spot} + s_2 ~~~ .     
\end{equation}
%%%%%%%%%%%%%%%%%%%%%%%%%
where $s_1 = 0.02  \pm 0.01 $ and  $s_2 =3.14 \pm 0.43$.

For the analysis of the plage area coverage we use the recent dataset published by
\citet{chatzistergos20}, that exhibits an increase in coverage levels during most minima compared to the dataset used in a previous reconstruction \citep{chatzistergos19}. Since the functional form in Eq. \ref{cycle_form} is by definition zero when $t = T0$, we prefer to use a power-law relationship between plage areas and sunspots, as suggested in several studies
\citep[see e.g.,][]{chapman1997, chapman2011,chatzistergos22}. We use the dataset of sunspot and plage area coverages within the common time range of 1903-2023 CE, and consider their monthly average. As suggested in \citet{chatzistergos22}, we find that the plage ($\alpha_f$) and the sunspot ($\alpha_s$) area are linked by a power-law relationship; in particular we find the following best fit:

%%%%%%%%%%%%%%%%%%%%%%%%
\begin{equation}
\label{plage_vs_spot}
\alpha_f = (1.3 \pm 0.1) \alpha_s^{(0.61 \pm 0.02)} +  (0.0042 \pm 0.0005)     
\end{equation}
%%%%%%%%%%%%%%%%%%%%%%%%%

We note that the parameters in Eq. \ref{plage_vs_spot} are different from those reported in \citet{chatzistergos22}. This is likely due to the use of monthly average values in our study instead of the daily values. We report that Spearman's rank correlation coefficient between sunspot and plage area is 0.930, statistically significant at $p < 0.01$.

%%%%%%%%%%%%%%%%%%%%%%%%%%%%%%%%%%%%%
\begin{figure}[h]
\centering
\includegraphics[scale=0.7]{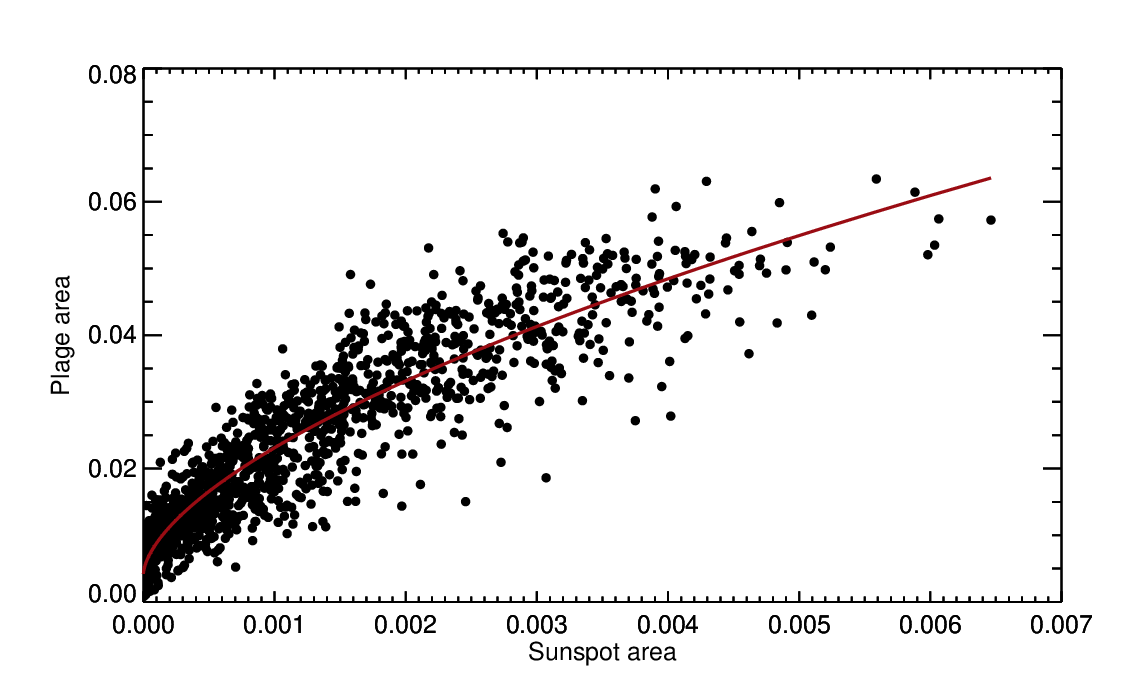}
\caption{Relationship between monthly plage and sunspot areas coverage for the years 1903 to 2023.
The red curve represents the power-law fit to the data (Eq. \ref{plage_vs_spot} in the text. The Spearman's rank correlation coefficient is 0.930, statistically significant at $p < 0.01$.}
\label{PLAGE_parametric}
\end{figure}
%%%%%%%%%%%%%%%%%%%%%%%%%%%%%%%%%%%%%%

Following the approach described in \citet{penza2022}, we investigate the correlation between the parameters in Eq. \ref{cycle_form} and the cycle-averaged $F_o$ values.
The values of $F_{o}$ are given with yearly cadence and substantial $\sigma$ error (20-40\%), so  we compute the average values through a Montecarlo simulation of one thousand iterations, where we calculate the average value over 11 years, varying the $F_{o}$ values within $3\sigma$. Then, we consider the mean of the averages with the corresponding error given by the standard deviation.
Given the high positive correlation  between 11-yr averaged values of $F_{o}$ ($\overline{F_{o}}$) and of sunspot area coverage, guaranteed by a Pearson correlation coefficient r = 0.81, with a confidence level greater than 95\% (p = 0.048), we search a correlation between $\overline{F_{o}}$ and the parameter $P_{k}$:
%%%%%%%%%%%%%%%%%%%%%%%%
\begin{equation}
\label{Pk_Fopen}
\nonumber P_{k} = a {\overline{F_{o}}} +  b    
\end{equation}
%%%%%%%%%%%%%%%%%%%%%%%%%
This is because we observe that the integral of the expression in Eq. \ref{cycle_form} over the cycle duration corresponds to the following expression
%%%%%%%%%%%%%%%%%%%%%
\begin{equation}
\label{cycle_int}
\frac{\int_{0}^{\tau_{k}} x_{k}(t)dt}{\tau_{k}}  =  \left(\frac{Td_{k}}{2Ts_{k}}\right)^{2} \left [\sqrt{\pi} \frac{\mathrm{erf}(f_{k})}{f_{k}} - 2 e^{-f_{k}^{2}} \right]
\end{equation}
%%%%%%%%%%%%%%%%%%%%%%%
where $f_{k} \equiv \tau_{k}/Td_{k}$ and erf is the so-called error function. From the latter expression, it follows that the average coverages are proportional to the parameter $P_{k}$. On the other hand, we have verified that the term $\left [\sqrt{\pi} \frac{\mathrm{erf}(f_{k})}{f_{k}} - 2 e^{-f_{k}^{2}} \right]$ in Eq. \ref{cycle_int} does not correlate with $\overline{F_{o}}$. 
%%%%%%%%%%%%%%%%%%%%%%%%% 
\begin{figure}[h]
\centering
\includegraphics[scale=0.7]{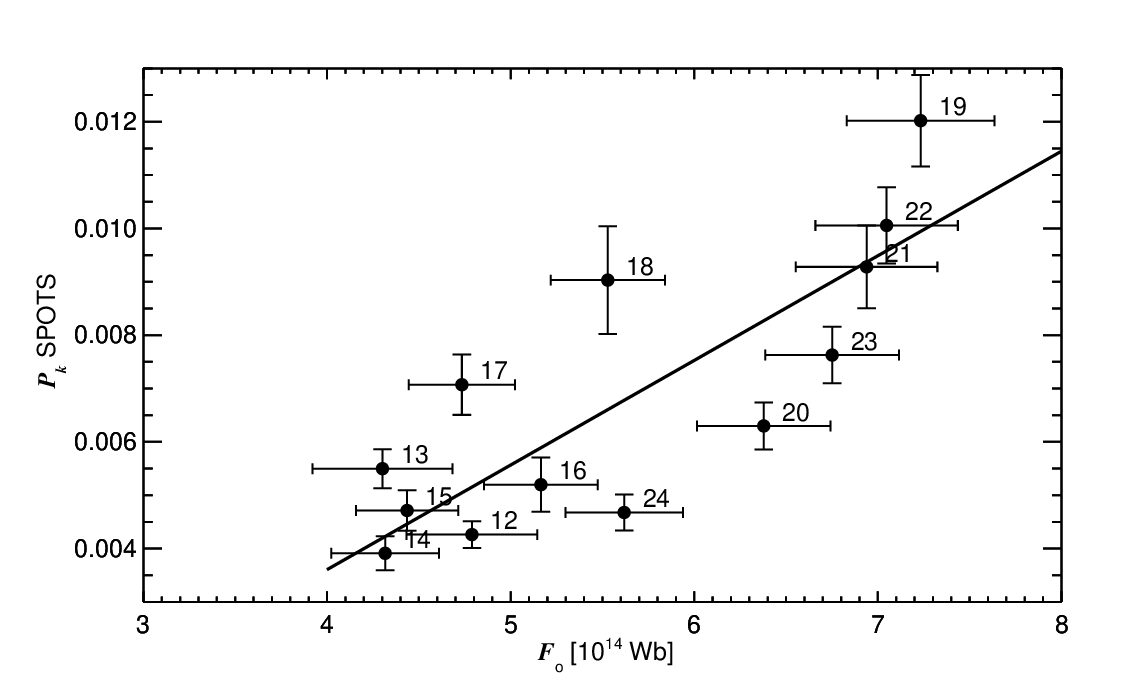}
\caption{Dependence of the spot parameter $P_k$ on the k-cycle averaged open solar magnetic flux $F_o$. The individual solar cycle numbers are indicated in the plot, with error bars representing the 1$\sigma$ standard error of the mean. This relationship is well described by the linear model (solid line) provided in Eq. 6. 
The Pearson's correlation coefficient is 0.80, with a confidence level greater than 95\% (p = 0.048).}
\label{Fopen_Pk}
\end{figure}
%%%%%%%%%%%%%%%%%%%%%%%%%%%%%%%%%%%%%%
By best fit of the Eq.\ref{Pk_Fopen} (Fig. \ref{Fopen_Pk}), we obtain the following values:
%%%%%%%%%%%%%%%%%%%%%%%%%%%%%%%%%%%%%%
\begin{eqnarray}
    \nonumber a = 0.002 \pm 0.001 \,\,\,\, (Wb^{-1}) \\
    \nonumber b = -0.004 \pm  0.003\\
\end{eqnarray}
%%%%%%%%%%%%%%%%%%%%%%%%%%%%%%%%%%

%%%%%%%%%%%%%%%%%%%%%%%%% 
\begin{figure}[h]
\centering
\includegraphics[scale=0.7]{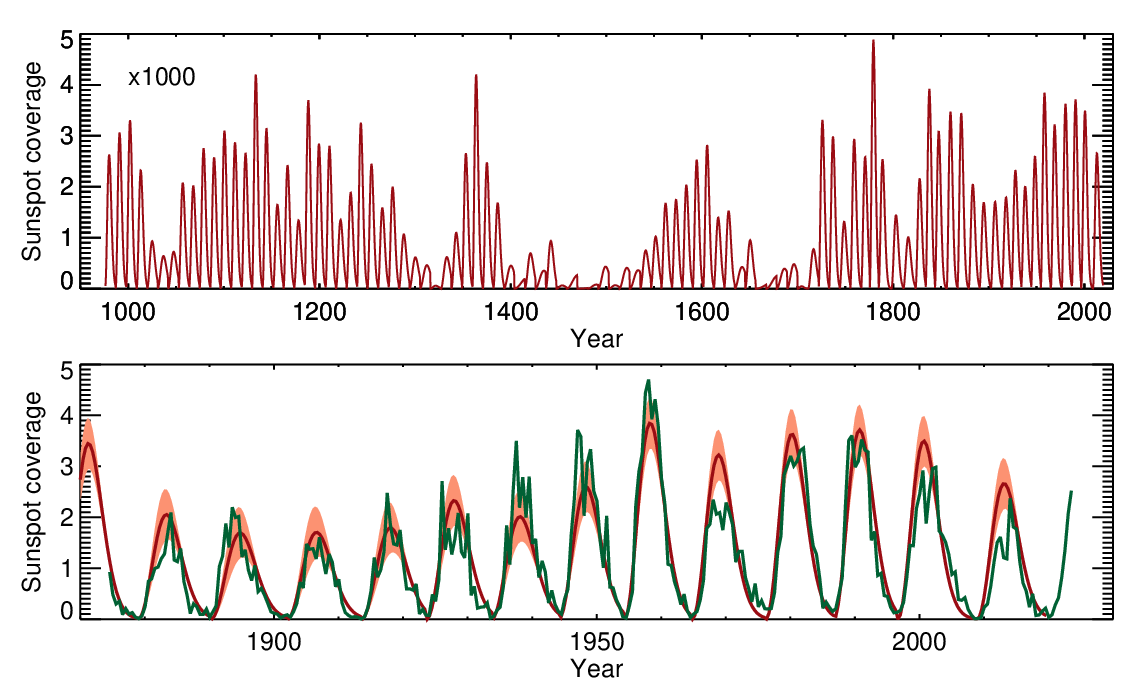}
\caption{Top panel: Reconstruction of sunspot area coverage from 971 CE to the present. Bottom Panel: The reconstructed time series (red curve) and its uncertainty range (light red region) are shown from 1900 to the present. The actual measured monthly sunspot area coverage by \citet{mandal2020} is shown in green.} 
\label{spot_rec}
\end{figure}
%%%%%%%%%%%%%%%%%%%%%%%%%%%%%%%%%%%%%%

\begin{figure}[h]
\centering
\includegraphics[scale=0.7]{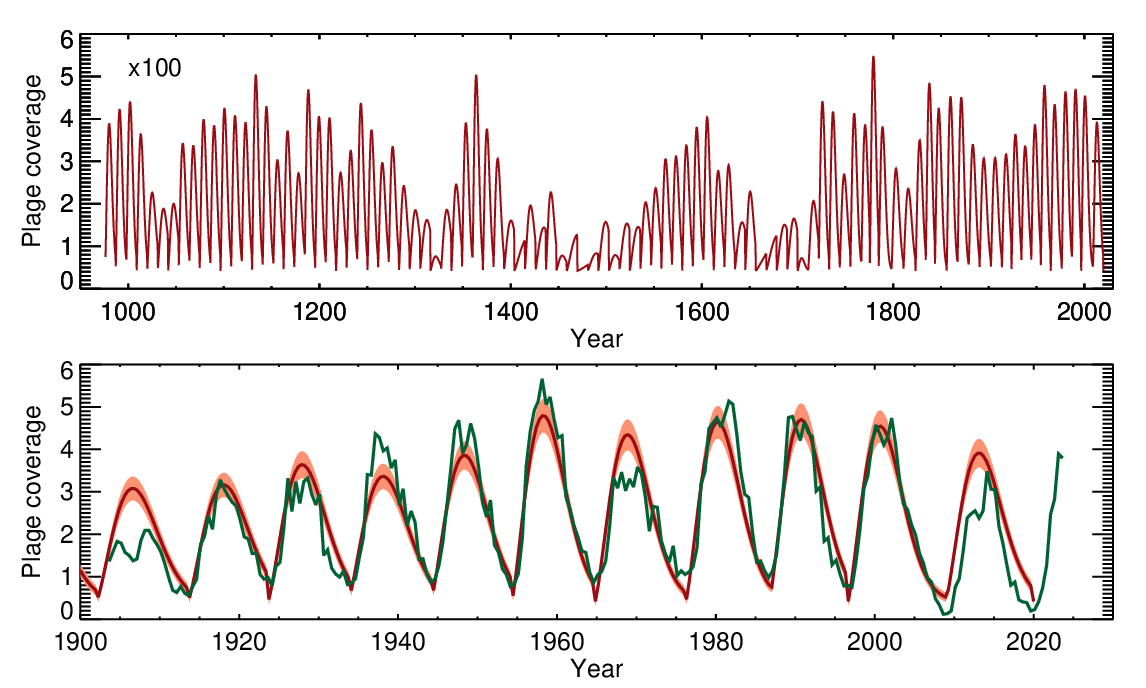}
\caption{Same as in Fig. \ref{spot_rec} but for the plage area coverage. The plage area composite by \citet{chatzistergos20} is shown in green.} 
\label{plage_rec}
\end{figure}
%%%%%%%%%%%%%%%%%%%%%%%%%%%%%%%%%%%%%%
Then we are able to obtain the values of the $P_{k}$ parameter for all of the past cycles, from 971 to 1900 CE. The $P_{k}$ values are obtained by application of this relation. We are now able to calculate the individual parameters $Ts_{k}$ and $Td_{k}$ from $P_{k}$ by using the relation provided in Eq. \ref{TsvsTd_spot}. Subsequently, we reconstruct the sunspot coverage trend (Fig. \ref{spot_rec}) by substituting these values into Eq. \ref{cycle_form}. Finally, the plage coverage is reconstructed using the relationship described in Eq. \ref{plage_vs_spot}. The resulting reconstruced plage area is shown in Fig. \ref{plage_rec}, together with the measured composite by \citet{chatzistergos20}.

\section{TSI reconstruction} 
\label{sec_tsi}
%%%%%%%%%%%%%%%%%%%%%%%%% 
\begin{figure}[h]
\centering
\includegraphics[scale=0.7] {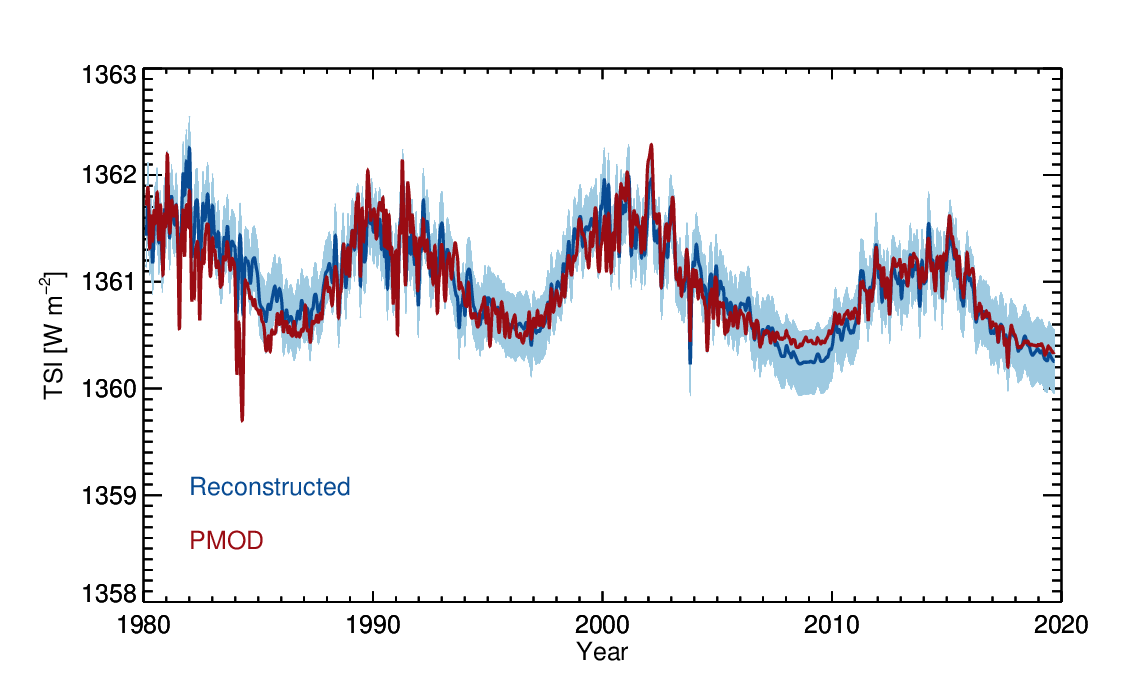}
\caption{Comparison between the TSI PMOD composite (red) and the reconstructed TSI time series (blue) from 1980 to 2020 CE (Eq. \ref{delta_TSI}). The light blue region represents the uncertainty range in the reconstructed time series.}
\label{tsi_pmod}
\end{figure}
%%%%%%%%%%%%%%%%%%%%%%%%%%%%%%%%
%%%%%%%%%%%%%%%%%%%%%%%%% 
\begin{figure}[h]
\centering
\includegraphics[scale=0.7] {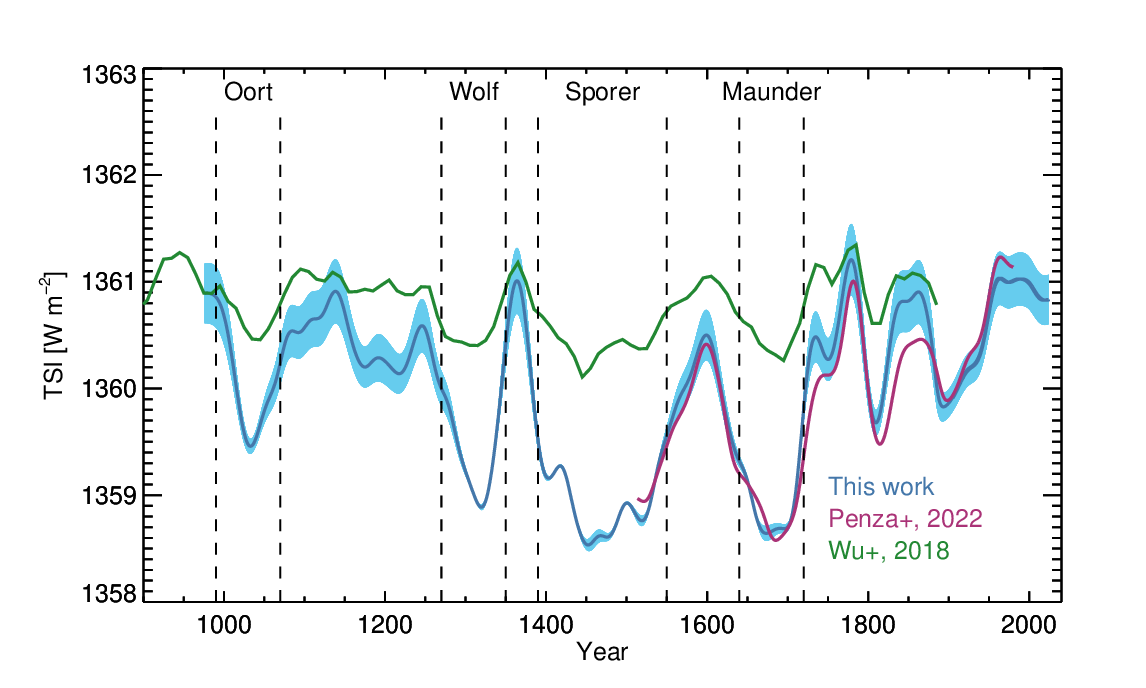}
\caption{Three different TSI reconstructions are shown for comparison. The blue line, covering the period from 971 to 2020 CE, along with its uncertainty range (light blue region), represents the results from this work. The TSI values from  \citet{wu2018} are depicted in green, while the purple line shows the reconstruction from our 2022 study \citet{penza2022}.}
\label{tsi_rec}
\end{figure}
%%%%%%%%%%%%%%%%%%%%%%%%%%%%%%%%
Solar irradiance variations can be accurately reproduced using a widely accepted method \citep[e.g.][]{penza2003, ermolli2011, fontenla2011, ball2014, Criscuoli2018, berrilli2020}. This method involves summing the fluxes generated by various solar components, each weighted according to its corresponding coverage area:
%%%%%%%%%%%%%%%%%%%%%
\begin{equation}
\label{Flux_rec}
F(t) =  \sum_{j} \alpha_{j}(t) F_{j},
\end{equation}
%%%%%%%%%%%%%%%%%%%%%%%
where $F_{j}$ is the integral over all wavelength of the spectrum of the flux from the j-feature (quiet, network, facula, and sunspot) - supposed time-independent - and $\alpha_{j}(t)$ is the respective coverage. 

In order to reconstruct the total solar irradiance variations, we proceed in the same way as in \citet{penza2022}, computing the TSI variations as: 
%%%%%%%%%%%%%%%%%%%%%%%%%%%%%%%%%%%%%%%%5
\begin{equation}
\label{Flux}
F(t) = \alpha_{f}(t) F_{f} +  \alpha_{n}(t) F_{n} + \alpha_{s}(t) F_{s} + (1 - \alpha_{f}(t) - \alpha_{n}(t) + \alpha_{s}(t)) F_{q}
\end{equation}
%%%%%%%%%%%%%%%%%%%%%%%
where the subscripts $f$, $n$, $s$ and $q$ indicate facular, network, sunspot and solar quiet contributions, respectively. Then, the relative variation has a very simple form:
%%%%%%%%%%%%%%%%%%%%%
\begin{equation}
\label{delta_Flux}
\delta F(t) \equiv \frac{F(t) - F_{q}}{F_{q}} = \alpha_{f}(t) \frac{F_{f} - F_{q}}{F_{q}} + \alpha_{n}(t)  \frac{F_{n} - F_{q}}{F_{q}} + \alpha_{s}(t)  \frac{F_{s} - F_{q}}{F_{q}} = \alpha_{f}(t) \delta_{f} + \alpha_{n}(t) \delta_{n} + \alpha_{s}(t) \delta_{s} 
\end{equation}
%%%%%%%%%%%%%%%%%%%%%%%
where $\delta_{f}$, $\delta_{n}$ and  $\delta_{s}$ are the relative contrasts. 

We adopt a linear relation between $\alpha_{f}(t)$ and $\alpha_{n}(t)$ \citep[see e.g.][]{Criscuoli2018, pooja2021}. 

As inferred by \citet{skumanich}, this relationship is the consequence of the presumed derivation of the network components from the decay of plage regions (see Fig.7 of \citet{skumanich}). By imposing the following relation:

%%%%%%%%%%%%%%%%%%%%%
\begin{equation}
 \label{an-af}
 \alpha_{n} = A_{n} +  B_{n} \alpha_{f}(t).
\end{equation}
%%%%%%%%%%%%%%%%%%%%%%% 

we can rewrite Eq. \ref{delta_Flux} as:

%%%%%%%%%%%%%%%%%%%%%
\begin{equation}
\label{delta_TSI}
\delta F(t) = C{n} + \alpha_{f}(t) \delta_{fn} + \alpha_{s}(t) \delta_{s}
\end{equation}
%%%%%%%%%%%%%%%%%%%%%%%

where $\delta_{fn} \equiv \delta_{f} + B_{n} \delta_{n}$ is the mixed facular and network contribution, while $C_{n} \equiv A_{n} \delta_{n}$ represents the value of the network contrast weighted by the corresponding fractional area at the solar minimum. 

The parameters $C_{n}$, $\delta_{fn}$ and $\delta{s}$ are obtained by fitting single cycles (Fig. \ref{tsi_pmod}), where we can neglect the long-term modulation, with the PMOD - TSI composite \footnote{\url{ftp://ftp.pmodwrc.ch/pub/data/irradiance/virgo/TSI/} (last accessed January 2023)} \citep{montillet22}. The values are slightly different with respect to \citet{penza2022}, as we have used a different plage coverage dataset and now we are fitting over three complete solar cycles (22, 23 and 24):

%%%%%%%%%%%%%%%%%%%%%%
\begin{eqnarray}
\label{tsi_parameter}
\nonumber \delta_{fn} = 0.041 \pm 0.002 \\ 
\nonumber \delta_{s} = -0.30  \pm  0.01\\ 
\nonumber C_{n} = 8.8 ^{.} 10^{-4} \pm   1 ^{.} 10^{-4}
\end{eqnarray}
%%%%%%%%%%%%%%%%%%%%%%%
Finally, we introduce a long-term modulation of the $C_{n}$ values

%%%%%%%%%%%%%%%%%%%%%
\begin{equation}
\label{delta_TSI2}
\delta F(t) = C_{n} \left[ 1 + F_{LT}(t) \right] N_{norm} + \alpha_{f}(t) \delta_{fn} + \alpha_{s}(t) \delta_{s}
\end{equation}
%%%%%%%%%%%%%%%%%%%%%%%

where $F_{LT}(t)$ is the normalized long-term modulation function computed in Sec. \ref{sec_emd} and $N_{norm}$ is a normalization parameter. 
We employ two free parameters, $N_{norm}$ and $F_{q}$, for the final reconstruction of F(t) (Eq. \ref{Flux}).
The best fit over the whole PMOD composite gives the values: $N_{norm} = 0.55$ and $F_{q} = 1359.83~W m^{-2}$, respectively.
The result is shown in Fig. \ref{tsi_rec}, where, for comparison, the reconstruction by \citet{penza2022} and that of \citet{wu2018} are included, with the latter only extending up to 1900 CE.

%--------------------------SECTION
\section{Discussion and Conclusions}
\label{sec_discussion}

In this work, we present a reconstruction of Total Solar Irradiance (TSI) variability over the last millennium, spanning from 971 to 2020 CE. This reconstruction builds upon the methodology proposed in our previous study \citep{penza2022}, where the Solar Modulation Potential ($\phi$) was used as the principal proxy for TSI variability. In the current study, we employ the Open Solar Magnetic Field ($F_{o}$) as the main proxy, allowing us to extend the reconstruction further back in time.  Additionally, this reconstruction incorporates updated values of plage coverages \citep{chatzistergos22} and TSI composites \citep{montillet22}.

Our approach integrates modern TSI measurements to estimate the contributions of sunspots and faculae in terms of their area coverages and bolometric contrasts. The resulting values for the bolometric contrasts reported in Sec.~\ref{sec_tsi} agree remarkably well with those reported in the literature \citep[e.g.][]{foukal2004, chapman1994, walton2003}. A third, long-term component of our model takes into account weak magnetic features not  associated with active regions \citep[e.g.][]{marchenko2022}, that modulates irradiance during minima.  This third component is estimated from the low-frequency terms obtained from Empirical Mode Decomposition (EMD) applied to $F_{o}$ and modern TSI measurements.

Our reconstruction (Fig.~\ref{tsi_rec}) reveals four significant periods of grand minima, including the Sp\"{o}rer and Maunder Minima. The Sp\"{o}rer Minimum, occurring between the 15th and 16th centuries, shows a TSI variation of approximately 2.3 $Wm^{-2}$ relative to present values, making it the deepest and longest minimum identified. We also estimate a growth in the TSI value of about 1 $W m^{-2}$ during the first half of last century. After the 1960s, this value has remained substantially constant (on average) until the beginning of this century.
The Maunder Minimum shows a variation of 2.2 $Wm^{-2}$ with respect to modern minima, which is very close to the value of 2.5 $Wm^{-2}$ obtained in \citet{penza2022}. The use of $F_{o}$ instead of $\phi$, the different plage coverages and TSI composites and the determination of the $F_{q}$ parameter by fit are the main difference between the reconstruction presented in this study and our previous work. 

The TSI variation between the Maunder Minimum and the present found our studies is higher than the values recently propose in the literature. For instance, SATIRE produces a 0.55 $Wm^{-2}$ variation \citep{wu2018}, in line with the more recent
estimate of 0.5 $Wm^{-2}$ by \citet{chatzistergos24}. The model by \cite{wang2021} produces a ~0.2 $Wm^{-2}$ variation. On the other hand, \citet{egorova2018} obtained a TSI change  
between 3.7 and 4.5 $Wm^{-2}$.

\citet{lockwood2020} placed a limit between 0.07 and -0.13 $Wm^{-2}$, indicating that, within the constraints imposed by modern measurements, the TSI during grand minima could have been even higher than during current minima.

Our TSI reconstruction, and its reliability, depend on the composite of direct TSI measurements used (PMOD) and the choice to use cosmogenic data as a proxy for long-term variations. On the other hand, all the past reconstructions presented in the literature are based on similar assumptions and are prone to uncertainties propagating from the observational data and data-products they rely on \citep[e.g.][]{egorova2018,lockwood2020,chatzistergos23}. Even models that try to tackle the issue from a different point of view, using 3D magneto-hydrodynamic (MHD) simulations \citep{yeo2020, chatzistergos23}, are inevitably bound by other assumptions, such as the small-scale dynamo not being affected by long-term variations. Despite the different approach, our result is compatible with the upper limit for TSI variations from the Maunder Minimum to the present (about 2 $Wm^{-2}$) estimated using MHD simulations by \cite{yeo2020}.

%--------------------------SECTION
\acknowledgments
The National Solar Observatory is operated by the Association of Universities for Research in Astronomy, Inc. (AURA), under cooperative agreement with the National Science Foundation. L.B. and S.C. are members of the international team on Reconstructing Solar and Heliospheric Magnetic Field Evolution Over the Past Century supported by the International Space Science Institute (ISSI), Bern, Switzerland.
This research was partially supported by the Italian MIUR-PRIN \emph{Circumterrestrial Environment: Impact of Sun-Earth Interaction\/} grant 2017APKP7T. M. Cantoresi and L. Lucaferri are supported by the Joint Research PhD Program in “Astronomy, Astrophysics and Space Science” between the University of Rome "Tor Vergata", the University of Rome "La Sapienza" and INAF. This research was part of the CAESAR (Comprehensive spAce wEather Studies for the ASPIS prototype Realization) project, supported by the Italian Space Agency and the National Institute of Astrophysics through the ASI-INAF n.2020-35-HH.0 agreement for the development of the ASPIS (ASI SPace weather InfraStructure) prototype of scientific data centre for Space Weather.

The authors gratefully thank the anonymous referee for helpful comments and suggestions, Ilya Usoskin for providing $F_{o}$ data, J\"{u}rg Beer for useful information about the cosmogenic-isotope measurements, and Valerio Formato for helpful discussions.

\bibliography{main}{}
\bibliographystyle{aasjournal}

\end{document}